%% file: PPPPP_apj.tex
\documentclass[twocolumn]{aastex702}

\usepackage{amsmath}
\providecommand{\degr}{\ensuremath{^\circ}}
\providecommand{\arcmin}{\ensuremath{^\prime}}

\providecommand{\farcs}{\mbox{$.\!\!^{\prime\prime}$}}

\shorttitle{Single pulses of Vela, PSR~J0437$-$4715, and PSR~J1644$-$4559}
\shortauthors{Lousto et al.}

\begin{document}

\title{Single-pulse reanalysis of the 2024 Vela glitch and new observations of
PSR~J0437$-$4715 and PSR~J1644$-$4559}

\author[0000-0002-6400-9640]{Carlos O. Lousto}
\affiliation{Center for Computational Relativity and Gravitation, Rochester
Institute of Technology, 85 Lomb Memorial Drive, Rochester, NY 14623, USA}
\email{colsma@rit.edu}

\author[0009-0007-0570-4196]{Ruby Shrestha}
\affiliation{Golisano College of Computing and Information Sciences, Rochester
Institute of Technology, Rochester, NY 14623, USA}
\email{rs9466@rit.edu}

\author[0000-0002-8027-0078]{Ezequiel Zubieta}
\affiliation{Instituto Argentino de Radioastronom\'ia (CCT La Plata, CONICET;
CICPBA; UNLP), C.C.5, (1894) Villa Elisa, Buenos Aires, Argentina}
\email{ezequielzubietaa1@gmail.com}

\author[0000-0003-4027-4826]{Susana B. Araujo Furlan}
\affiliation{Instituto Argentino de Radioastronom\'ia (CCT La Plata, CONICET;
CICPBA; UNLP), C.C.5, (1894) Villa Elisa, Buenos Aires, Argentina}
\affiliation{Facultad de Matem\'atica, Astronom\'ia, F\'isica y Computaci\'on,
UNC, Av. Medina Allende s/n, Ciudad Universitaria, CP:X5000HUA C\'ordoba,
Argentina}
\email{susana.araujo@unc.edu.ar}

\author[0000-0003-1282-3031]{Guillermo Gancio}
\affiliation{Instituto Argentino de Radioastronom\'ia (CCT La Plata, CONICET;
CICPBA; UNLP), C.C.5, (1894) Villa Elisa, Buenos Aires, Argentina}
\email{ganciogm@gmail.com}

\author[0000-0001-9072-4069]{Federico Garc\'ia}
\affiliation{Instituto Argentino de Radioastronom\'ia (CCT La Plata, CONICET;
CICPBA; UNLP), C.C.5, (1894) Villa Elisa, Buenos Aires, Argentina}
\affiliation{Facultad de Ciencias Astron\'omicas y Geof\'isicas, Universidad
Nacional de La Plata, Paseo del Bosque, B1900FWA La Plata, Argentina}
\email{fgarcia1012002@gmail.com}

\author[0000-0002-1566-9044]{Santiago del Palacio}
\affiliation{Instituto Argentino de Radioastronom\'ia (CCT La Plata, CONICET;
CICPBA; UNLP), C.C.5, (1894) Villa Elisa, Buenos Aires, Argentina}
\affiliation{Department of Space, Earth and Environment, Chalmers University of
Technology, SE-412 96 Gothenburg, Sweden}
\email{santiagodp1990@gmail.com}

\author[0000-0002-9509-6775]{Ryan Missel}
\affiliation{Golisano College of Computing and Information Sciences, Rochester
Institute of Technology, Rochester, NY 14623, USA}
\email{rxm7244@rit.edu}

\author[0000-0002-5678-2369]{Linwei Wang}
\affiliation{Golisano College of Computing and Information Sciences, Rochester
Institute of Technology, Rochester, NY 14623, USA}
\email{lxwast@rit.edu}

\correspondingauthor{Carlos O. Lousto (colsma@rit.edu)}

\begin{abstract}
The Pulsar Monitoring in Argentina (PuMA) collaboration systematically
monitors southern glitching pulsars, maintaining high-cadence single-pulse
records of the Vela pulsar. We present a pulse-per-pulse reanalysis of the
2024 major glitch of Vela (PSR~J0835$-$4510) with the 400~MHz-bandwidth ROACH
backend of the Argentine Institute of Radioastronomy, and extend our
machine-learning single-pulse pipeline---Isolation Forest outlier rejection,
$\beta$-Variational-AutoEncoder denoising, and Self-Organizing-Map
clustering---to new observations of PSR~J1644$-$4559 and the millisecond
pulsar PSR~J0437$-$4715. For Vela, the 4- and 6-cluster decompositions of the
eight days bracketing the glitch reproduce, with seven times the previous
bandwidth, the behavior found with the narrow-band ETTUS receivers:
higher-amplitude clusters peak earlier in phase, are narrower, more skewed,
and less populated. With the glitch jump and its two exponential recovery
terms included in the timing solution, the mean profile is stable to 1\%
across all eight days (width change $-0.5\pm0.9$\% from pre- to post-glitch);
omitting the recovery terms would mimic a post-glitch broadening of up to
55\% through a folding-frequency error at the $10^{-7}$ level. The clusters
of PSR~J1644$-$4559 differ almost exclusively in amplitude, as expected for a
scattering-dominated profile. For PSR~J0437$-$4715, retaining the 10\% of
pulses with the highest peak-dominance score doubles the signal-to-noise
ratio, and a five-cluster decomposition yields narrow, phase-ordered groups a
factor $3.6\pm0.4$ narrower than the average profile, suggesting a
$\sim$3.5-fold improvement in cluster-based timing precision for this
pulsar-timing-array target, to be confirmed in a follow-up paper.
\end{abstract}


\section{Introduction}

Pulsars are highly magnetized, rapidly rotating neutron stars emitting collimated beams of electromagnetic radiation from their magnetic poles. Although their rotation is extraordinarily stable, allowing them to act as highly precise astrophysical clocks, a subset of (mostly young) pulsars undergo spontaneous spin-up events known as glitches. These abrupt increases in rotational frequency are generally understood to be triggered by the transfer of angular momentum from a rapidly rotating superfluid interior to the solid crust of the star \citep{Anderson1975,Haskell2015}. Observations of glitches and their subsequent recoveries are therefore crucial probes of the dense-matter equation of state and of the internal structure of neutron stars.

The Vela pulsar (PSR~J0835$-$4510) is one of the most prolific glitching pulsars, typically undergoing a major (giant) glitch every two to three years. Because Vela is extraordinarily bright, it presents an ideal laboratory for single-pulse studies, yielding tens of thousands of pulses in a standard multi-hour observation. Analyzing the variations of these individual pulses provides unique insights into the varying structure of the emission regions within the pulsar's magnetosphere.

With the drastic increase in the volume of data generated by modern radio telescopes, machine learning has emerged as an indispensable tool for pulsar studies. Our collaboration recently implemented a novel unsupervised machine-learning pipeline to study Vela's individual pulses using the two radio telescopes of the Argentine Institute of Radioastronomy \citep[IAR;][]{Lousto:2021dia}. By applying Density-Based Spatial Clustering of Applications with Noise (DBSCAN), we associated pulses primarily by amplitude, confirming a correlation between higher-amplitude pulses and earlier arrival times \citep{Lousto:2021dia}. Using a Variational AutoEncoder (VAE) to reconstruct single pulses out of the background noise, together with Self-Organizing Map (SOM) clustering, we successfully modeled pulse variations indicative of emission regions located at stratified heights within the magnetosphere \citep{Lousto:2021dia}.

The capabilities of this automated single-pulse pipeline naturally extend to the analysis of transient phenomena such as glitches. The unpredictable nature of glitches makes real-time, pulse-per-pulse observations exceedingly rare, yet highly informative regarding the coupled dynamics between the neutron-star interior and the magnetosphere. Through the continuous high-cadence monitoring program at IAR, we successfully captured data surrounding the major Vela glitch of 2021 July 22 \citep{Zubieta:2022umm}. While precision timing provided insight into the fractional frequency jump and its corresponding decay timescales, our implementation of the VAE--SOM pipeline indicated no major qualitative systematic changes in the pulse cluster distributions during the days directly surrounding the event \citep{Zubieta:2022umm}.

Most recently, our campaign monitored the subsequent major Vela glitch, which occurred on 2024 April 29 \citep{Zubieta:2025rud}. The initial rotational characterization, derived from the surrounding days of observations, precisely determined two post-glitch recovery timescales of approximately 3 and 17 days \citep{Zubieta:2025rud}. VAE reconstructions during this period further quantified a correlation whereby higher amplitudes manifest with narrower pulse widths; however, the day-to-day clustering analysis once again demonstrated an overall qualitative stability of the magnetospheric emission across the glitch epoch \citep{Zubieta:2025rud}. This particular event has since acquired a multimessenger dimension: it was the target of the first dedicated search for gravitational waves associated with a Vela glitch, performed on data from the fourth LIGO--Virgo--KAGRA (LVK) observing run \citep{LIGOScientific:2025dup}.

To investigate whether the magnetospheric alterations accompanying a glitch occur on much shorter or subtler timescales than previously resolved, we present here a deeper, pulse-per-pulse reanalysis of the 2024 Vela glitch, based on the wider-bandwidth ROACH backend and on a pipeline upgraded with an explicit outlier-rejection stage and a $\beta$-VAE reconstruction. Building on our VAE and SOM architectures, this work aims to trace micro-variations immediately leading up to, and recovering from, the glitch epoch. Furthermore, to demonstrate the generalizability of our clustering techniques beyond Vela, we expand our study to report new single-pulse analyses of the millisecond pulsar PSR~J0437$-$4715 and of the long-period pulsar PSR~J1644$-$4559, investigating their distinct clustering topologies.

The paper is organized as follows. In Section~\ref{sec:observations} we describe the observations and the scientific motivation for single-pulse studies of each of the three pulsars. In Section~\ref{sec:techniques} we review the IAR instrumentation, the new ROACH-based digital backend, and the updates to our machine-learning pipeline. Section~\ref{sec:results} presents the results for each pulsar, and Section~\ref{sec:conclusions} contains our conclusions and a discussion of follow-up studies. Detailed tables of the clustering analyses are collected in the appendices.

\section{Observations}\label{sec:observations}

All the observations reported in this paper were carried out with the second of the two 30-m antennas of the IAR, located near La Plata, Argentina: A2 (``Esteban Bajaja''), the first one being A1 (``Carlos M. Varsavsky''). While our previous single-pulse analyses of Vela \citep{Lousto:2021dia,Zubieta:2022umm,Zubieta:2025rud} were based on data acquired with the original SDR-based (ETTUS) digital backends with 56~MHz of bandwidth per board \citep{Gancio:2019frj}, the present work makes use of the new ROACH-based digital receiver installed at A2 (hereafter R2; see Section~\ref{sec:techniques}), which provides a much wider instantaneous bandwidth of 400~MHz in dual circular polarization and an improvement of up to a factor of 2.5 in the signal-to-noise ratio (S/N) \citep{guilleRevistaMexicana}. These improvements are particularly important for extending single-pulse studies to fainter and faster pulsars, such as the millisecond pulsar PSR~J0437$-$4715, and to strongly scattered pulsars, such as PSR~J1644$-$4559 \citep[see the discussion in][]{Lousto:2023yvu}. In the following subsections we briefly review the scientific interest of single-pulse studies for each of the three targets.

The observational setup adopted for each of the three pulsars is summarized in
Table~\ref{tab:obslog}. The sampling interval was matched to the rotation period
of each target, from 327.68~$\mu$s for the 455~ms period of PSR~J1644$-$4559 down
to 20.48~$\mu$s for the 5.76~ms period of PSR~J0437$-$4715, so that the pulse
profile is always resolved by a few hundred phase bins. A typical observation
therefore yields $\sim1.5\times10^{5}$ single pulses for Vela,
$\sim2.8\times10^{4}$ for PSR~J1644$-$4559, and $\sim1.2\times10^{6}$ for
PSR~J0437$-$4715.

\begin{deluxetable*}{lccccccc}
\tablecaption{Setup of the R2 single-pulse observations analyzed in this work.\label{tab:obslog}}
\tablewidth{0pt}
\tablehead{\colhead{Pulsar} & \colhead{Epochs} & \colhead{$N_\mathrm{obs}$} &
\colhead{$\nu_\mathrm{c}$ (MHz)} & \colhead{Bandwidth (MHz)} &
\colhead{$t_\mathrm{samp}$ ($\mu$s)} & \colhead{$T_\mathrm{obs}$ (h)} &
\colhead{$N_\mathrm{pulses}$}}
\startdata
PSR~J0835$-$4510 & 2024 Apr 17--20, May 1--4 & 8 & 1400.4 & 400 & 163.84 & $\sim$3.6 & $\sim1.5\times10^{5}$ \\
PSR~J1644$-$4559 & 2026 Mar 3--5             & 3 & 1400.4 & 400 & 327.68 & $\sim$3.5 & $\sim2.8\times10^{4}$ \\
PSR~J0437$-$4715 & 2026 May 5, 18, 21        & 3 & 1425.0 & 351 & 20.48  & $\sim$2.0 & $\sim1.2\times10^{6}$
\enddata
\tablecomments{The 400~MHz band is divided into 512 channels of 0.78125~MHz each
(1200.8--1600~MHz); for PSR~J0437$-$4715 the lowest 63 channels were discarded,
leaving 449 channels over 1250--1600~MHz. Both circular polarizations are summed.
$T_\mathrm{obs}$ and $N_\mathrm{pulses}$ are typical values for a single
observation.}
\end{deluxetable*}

\subsection{Relevance of studying individual pulses of PSR~J0835$-$4510 (Vela)}

Historically, pulsar astronomy has relied heavily on integrated profiles---stable, average pulse shapes produced by folding thousands of individual radio pulses. While this technique is fundamental for precision timing, it inherently averages out the highly dynamic, underlying physics of the pulsar's magnetospheric emission. Because PSR~J0835$-$4510 (the Vela pulsar) is exceptionally bright, it constitutes an ideal, naturally occurring laboratory for probing high-resolution magnetospheric physics that would otherwise be lost in the background noise of standard receivers \citep{Krishnan2019,Lousto:2021dia}.

Studying the individual, un-averaged pulses of Vela provides direct observational constraints on the physics of plasma generation and coherent radio emission mechanisms. Individual pulses exhibit extreme pulse-to-pulse variability in amplitude, phase, and shape, and analyzing these variations helps to accurately map emission altitudes. For instance, the strong correlations between pulse amplitude and arrival phase observed in Vela suggest that different pulse emission modes originate from stratified heights within the open magnetic field lines \citep{Lousto:2021dia,Zubieta:2025rud}.

Furthermore, Vela is renowned for undergoing major spin-up events, or glitches, typically every two to three years \citep{Radhakrishnan1969}. While glitches are primarily driven by internal neutron-star dynamics, continuous single-pulse monitoring allows us to investigate the immediate, short-timescale response of the external magnetosphere to these violent internal events. Notably, high-cadence observations of the 2016 Vela glitch revealed transient nulling and changes in the single-pulse emission profile precisely at the epoch of the spin-up \citep{Palfreyman2018}. Detecting shifts in the single-pulse clustering or amplitude distributions directly surrounding a glitch epoch can reveal transient alterations of the magnetic field structure or of the particle acceleration zones \citep{Zubieta:2022umm}.

Finally, the sheer volume of data generated by continuous single-pulse observations of Vela serves as a rigorous testing ground for advanced statistical and analytical methods. A typical multi-hour observation yields tens of thousands of distinct pulses, necessitating modern unsupervised machine-learning techniques to efficiently separate signal from noise and to cluster distinct emission modes \citep{Lousto:2021dia}. The automated pipelines refined on the robust data sets of PSR~J0835$-$4510 establish a crucial baseline for tracking magnetospheric stability across major glitches \citep{Zubieta:2022umm,Zubieta:2025rud}, and can subsequently be applied to fainter, more complex transient phenomena.

\subsection{Relevance of studying individual pulses of PSR~J1644$-$4559}

In addition to the extensive monitoring of the Vela pulsar, extending our single-pulse machine-learning pipeline to PSR~J1644$-$4559 (also known as B1641$-$45) provides a highly valuable comparative study. PSR~J1644$-$4559 is a bright, long-period southern pulsar ($P \approx 0.455$~s) and, much like Vela, is a well-known glitching pulsar that undergoes frequent timing irregularities \citep{Zubieta:2024ynv}. Analyzing its un-averaged radio emission allows us to test the robustness and generalizability of our VAE and SOM architectures on a target with a distinctly different magnetospheric topology and rotational dynamics.

Historically, single-pulse studies of PSR~J1644$-$4559 have revealed a complex emission phenomenology, including log-normal amplitude distributions, distinct sub-pulse features, and orthogonal polarization modes that occasionally disrupt the smooth position-angle swing \citep{Karastergiou2004}. Furthermore, testing our clustering algorithms on this pulsar investigates whether the amplitude-dependent phase shifting and magnetospheric stratification observed in Vela \citep{Lousto:2021dia} are universal traits among bright glitching pulsars, or whether they are unique to Vela's specific emission geometry.

A uniquely compelling scientific driver for studying the high-cadence individual pulses of PSR~J1644$-$4559 is its complex interaction with the interstellar medium (ISM). Because of its high dispersion measure, the emission of PSR~J1644$-$4559 is significantly scatter-broadened by intervening plasma. Remarkably, this pulsar represents the only known instance of natural, fast-switching stimulated emission (an interstellar maser) pumped by pulsar photons, initially discovered in OH transitions \citep{Weisberg2005}. More recent studies have used the pulsar's line of sight to probe turbulence-induced tiny-scale atomic structures (TSAS) within the cold neutral medium \citep{Liu2025}.

While integrated profiles provide an average measure of these scattering tails and ISM interactions, single-pulse observations provide a natural differential tracer. Monitoring variations in the un-averaged pulses allows for precise measurements of pulse-to-pulse fluctuations in the scattering timescale, tracing turbulence and plasma structures on micro-astronomical-unit scales. Ultimately, extending our automated single-pulse analyses to targets such as PSR~J1644$-$4559 demonstrates the capability of the IAR pipeline to classify complex, scatter-broadened, and highly variable emission mechanisms beyond the optimal, unscattered conditions of the Vela pulsar.

\subsection{Relevance of studying individual pulses of PSR~J0437$-$4715}

To further demonstrate the versatility of our unsupervised machine-learning pipeline across vastly different evolutionary stages of neutron stars, we include a new single-pulse analysis of the millisecond pulsar (MSP) PSR~J0437$-$4715. Discovered as the closest and brightest known MSP \citep{Johnston1993}, PSR~J0437$-$4715 resides in a binary system with a white-dwarf companion. While Vela and PSR~J1644$-$4559 are young(er), slowly rotating pulsars with strong surface magnetic fields, PSR~J0437$-$4715 is a fully recycled pulsar characterized by a rapid rotational period ($P \approx 5.75$~ms) and a comparatively weak magnetic field.

The primary challenge in studying the un-averaged emission of MSPs is their inherently low flux per single rotation, which typically buries individual pulses within the instrumental noise floor. Thanks to its extreme proximity and brightness, PSR~J0437$-$4715 is one of the very few MSPs whose single-pulse emission can be resolved with high fidelity \citep{Ables1997,Vivekanand1998}. Applying our VAE and SOM architectures to this source allows us to probe whether the magnetospheric clustering topologies and amplitude-dependent variations observed in young pulsars persist in the compact, rapidly rotating magnetospheres of recycled pulsars.

Furthermore, studying the individual pulses of PSR~J0437$-$4715 holds critical importance for gravitational-wave astronomy. This pulsar is a cornerstone target for the global Pulsar Timing Arrays (PTAs) aiming to characterize the nanohertz gravitational-wave background. The ultimate limit to the timing precision of bright MSPs like PSR~J0437$-$4715 is ``phase jitter''---intrinsic, pulse-to-pulse variations in the emission phase and shape \citep{Oslowski2011,Shannon2014}.

By systematically mapping the single-pulse variability of PSR~J0437$-$4715 with our clustering algorithms, we aim to better characterize the statistical distributions underlying this jitter. Identifying specific, distinct subpopulations of single pulses within its complex integrated profile could eventually provide pathways for jitter mitigation, thereby enhancing the ultimate timing precision achievable for this key PTA pulsar. Extending our automated pipeline from bright, slow pulsars to a rapid, stable MSP fundamentally validates the generalizability of our machine-learning approach.

\section{Updated techniques}\label{sec:techniques}

\subsection{IAR instrumentation and the ROACH-based backend}\label{sec:IAR}

The IAR observatory, located at latitude $-34\degr 51\arcmin 57\farcs35$ and longitude $-58\degr 08\arcmin 25\farcs04$, within the Pereyra Iraola Park near the city of La Plata, Argentina, operates two 30-m single-dish antennas, A1 and A2, aligned in the North--South direction and separated by 120~m. They cover a declination range of $-90\degr < \delta < -10\degr$ and an hour-angle range of two hours east/west, allowing daily observations of a given source for up to $\sim$4 hours in the 1400~MHz band (L~band). A thorough description of the front ends and of the initial digital backends, based on ETTUS software-defined-radio boards providing 56~MHz of instantaneous bandwidth per board, is given in \citet{Gancio:2019frj}, together with an analysis of the radio-frequency-interference (RFI) environment showing that the 1--2~GHz band has a low level of RFI activity, adequate for radio astronomy, despite the IAR not being located in an RFI-quiet zone. This configuration enabled the systematic pulsar monitoring program of the PuMA collaboration \citep{Zubieta:2022umm,Zubieta:2024ynv} and our previous single-pulse studies of Vela \citep[see also the review in][]{Lousto:2023yvu}.

In mid-2022, a parallel digitizer system based on ROACH (Reconfigurable Open Architecture Computing Hardware) boards \citep{hickish2016casper} was added to both antennas\footnote{\url{https://casper.astro.berkeley.edu/wiki/ROACH}}$^{,}$\footnote{\url{https://digicom.org/roach-board.html}}. The ROACH-1 backends are configured to observe in dual circular polarization over a bandwidth of 400~MHz centered at 1400~MHz, with integration times down to 41 or 82~$\mu$s, providing an improvement of up to a factor of 2.5 in S/N with respect to the ETTUS boards \citep{guilleRevistaMexicana}. A flux calibration of A2, based on a noise diode injected into the front end, is described in \citet{calibDiodoDeRuido}; we do not apply it in the present analysis, and all single-pulse amplitudes in this paper are therefore reported in normalized (arbitrary) units. Throughout this paper we denote by R2 the ROACH backend installed at antenna A2, with which all the observations analyzed here were acquired; note that our previous single-pulse papers were instead based on A1/A2 ETTUS data. It is this increased sensitivity per pulse that makes single-pulse studies of the millisecond pulsar PSR~J0437$-$4715 and of the scatter-broadened pulsar PSR~J1644$-$4559 feasible at IAR, as anticipated in \citet{Lousto:2023yvu}.

For the single-pulse analyses presented here, the filterbank data are first
cleaned of radio-frequency interference in two stages: \texttt{RFIClean}
\citep{Maan:2020ent}, which excises periodic interference in the Fourier domain,
is run with the rotation frequency of the pulsar supplied explicitly so that its
own periodic signal is preserved, and a further mask is then derived with the
\texttt{rfifind} task of \texttt{PRESTO} \citep{presto}. The data are then incoherently dedispersed with \texttt{PRESTO}
and \texttt{PSRCHIVE} \citep{hotan_psrchive_2004}, and the resulting time series
is cut into individual rotations using the instantaneous topocentric period
predicted by the corresponding \texttt{polyco} file, which is re-evaluated every
100 pulses. For PSR~J0835$-$4510 the underlying timing solution is the full
glitch solution derived from the IAR monitoring campaign in \citet{Zubieta:2025rud}
(their Table~1), referred to the epoch MJD~60408 with
$\nu=11.18285953$~Hz and $\dot\nu=-1.55405\times10^{-11}$~Hz~s$^{-1}$, and
including the glitch at MJD~60429.86961 with its permanent frequency and
spin-down jumps and two exponential recovery terms with decay timescales of
17.3 and 2.78 days. For the observations of PSR~J0835$-$4510 discussed below,
the \texttt{polyco} models are tabulated in 60-min blocks with 12 coefficients
each, the epoch-dependent behavior---including the glitch and its
relaxation---being carried by the polynomial coefficients. Across the eight days
analyzed they give instantaneous topocentric periods between 89.4259 and
89.4265~ms. Since the post-glitch relaxation is explicitly modeled, the residual
error of the folding frequency is negligible on all days, a point we return to
in Sections~\ref{sec:J08} and~\ref{sec:conclusions}.
The individual pulses are sampled onto a
grid of $N_\mathrm{bin}=545$, 1388, and 281 phase bins per period for
PSR~J0835$-$4510, PSR~J1644$-$4559, and PSR~J0437$-$4715, respectively, the bin
width being the sampling interval itself (163.84, 327.68, and 20.48~$\mu$s), so
that the grid covers all but the last $\lesssim$0.15\% of each rotation.
The residual intra-channel dispersion smearing,
$\sim$160~$\mu$s for Vela and $\sim$6~$\mu$s for PSR~J0437$-$4715, is in both cases
at or below the sampling interval; for PSR~J1644$-$4559 it amounts to
$\sim$1.1~ms, still well below the several-millisecond scattering tail that
dominates the observed width of its profile.

\subsection{Outlier detection with Isolation Forests}\label{sec:outliers}

A pulsar single-pulse data set may contain a few non-uniform signals that appear visibly different from the general signal structure, typically produced by RFI or instrumental dropouts. We refer to such non-uniform signals as outliers. We observed that the presence of outliers distorts the SOM output, leading to some clusters exhibiting atypical mean signal structures and thereby compromising the analysis.
We therefore employ the Isolation Forest algorithm \citep{liu2008isolation} to filter out outliers from the data set prior to performing the VAE and SOM analyses.

Isolation Forest is an unsupervised anomaly-detection algorithm based on the intuition that outliers are comparatively easier to isolate than typical samples, and hence have shorter isolation paths. An ensemble of random isolation trees is built by recursively splitting the data at randomly chosen values of selected features until each point is isolated (or a predefined depth is reached). The average isolation-path length of each sample across the ensemble is converted into an anomaly score, and a threshold on this score determines the outlier list \citep[for the precise definitions, see][]{liu2008isolation}. We employ the Isolation Forest implementation of \texttt{scikit-learn} \citep{pedregosa2011scikit}\footnote{\url{https://scikit-learn.org/stable/modules/generated/sklearn.ensemble.IsolationForest.html}} with default hyperparameters.

Our general application methodology proceeds as follows: (i) we determine the pulse center as the position of the maximum of the average pulse; (ii) we window each pulse to $\pm 100$ bins around the mean pulse center, resulting in pulses of length 200 bins; (iii) we rebase each pulse by subtracting its mean; (iv) we run a $10\times10$ SOM on the signals, producing 100 clusters; (v) we apply the Isolation Forest algorithm to the mean signals of these 100 clusters, using a set of features tailored to the outlier detection, in order to identify the outlier clusters; (vi) among these, we retain as outliers only the clusters whose population does not exceed a count threshold; and (vii) we remove the pulses belonging to the selected outlier clusters from the data set. 

We introduce the cluster-count threshold to prevent overly aggressive outlier removal and the consequent exclusion of a substantial number of pulses. Upon careful analysis, we observed that a few clusters that were flagged as outliers exhibited only minor deviations from the general pattern with respect to the predefined feature set, while retaining mean-pulse morphologies broadly consistent with those of valid clusters and containing a relatively large proportion of pulses. Considering this, for each pulsar, we define the count threshold empirically through visual inspection of outlier SOM clusters across all selected observation days, setting it based on the count of the largest outlier cluster exhibiting clear morphological inconsistencies. We present the methodology customized for each pulsar as follows.

\subsubsection{PSR~J0835$-$4510} We use the general methodology as described, with the following set of features in step (v): the size of the cluster (number of pulses); the $R^2$ score (fraction of variance explained by the mean cluster signal); the position of the peak; the range of the mean cluster signal; and the slope of the mean cluster signal. The count threshold in step (vi) is set to 0.01\% of the total number of pulses.

\subsubsection{PSR~J1644$-$4559} For this pulsar we noted that the pulse baselines did not consistently align at zero, even after the initial rebasing of step (iii). We therefore use a modified feature set designed to focus on the approximate baseline position: we omit the windowing steps (i)--(ii)---considering the entire pulse phase so that the peak does not over-influence the fit---and in step (v) we flag as outliers the clusters whose mean signal has a best-fit-line $y$-intercept $\geq 0.5$. The count threshold in step (vi) is set to 0.25\% of the total number of pulses.

\subsubsection{PSR~J0437$-$4715} For this much fainter (per pulse) data set, in which a dominant peak appears in only a minor fraction of the single pulses, we first implement a filtering stage designed to retain only the pulses that display a distinct peak, and then apply the Isolation Forest algorithm to the filtered data set following the general methodology with one difference: we consider a $20\times20$ SOM instead of a $10\times10$ SOM for PSR~J0437$-$4715 since the former led to more visibility of the outliers and eased their detection. The count threshold in step (vi) is set to 0.05\% of the total number of pulses.

\begin{figure*}[!ht]
\centering
\includegraphics[width=\textwidth]{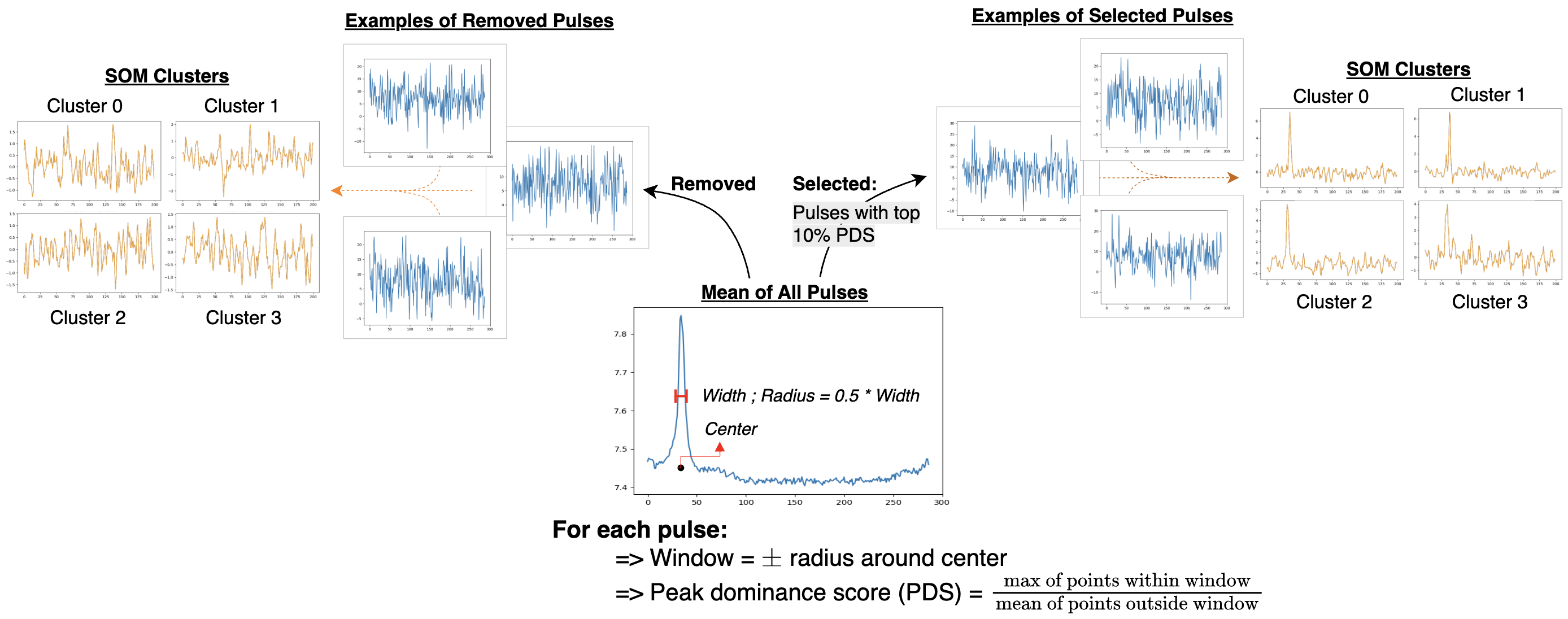}
 \caption{Visual representation of the filtering process for PSR~J0437$-$4715.}
\label{fig:J04_filtering_process}
\end{figure*}

\begin{figure}[!ht]
   \centering
   \includegraphics[width=\columnwidth]{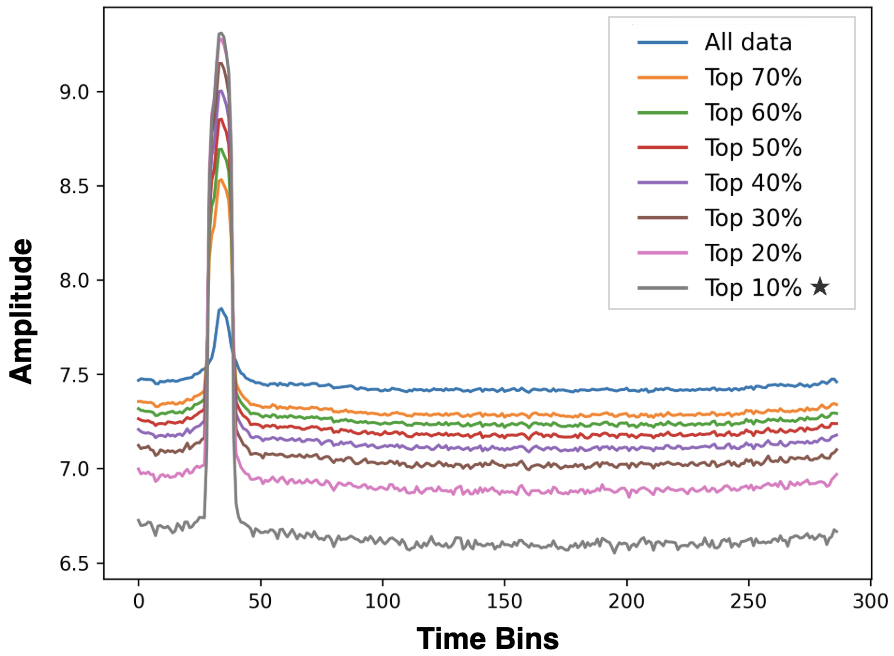}
     \caption{Mean pulses for the complete data set and for subsets corresponding to the top 70\% to top 10\% of peak-dominance scores.}
  \label{fig:J04_filtering_threshold}
 \end{figure}

Our filtering stage is illustrated in Figure~\ref{fig:J04_filtering_process}. It proceeds as follows: (i) we determine the mean pulse of the entire data set, its center (the position of the maximum of the mean pulse), and its peak width (computed as the full width at half maximum around the peak of the mean pulse); (ii) we define the radius as half the peak width of the mean pulse, and for each pulse, consider the window as $\pm \text{radius}$ bins around the mean pulse center; (iii) we compute each pulse's peak-dominance score (PDS) as the ratio of the maximum value within the window to the mean value outside the window; (iv) we retain only the pulses whose peak-dominance score falls in the top 10\%, which offers a higher signal-to-noise ratio and a better baseline (as shown in Figure~\ref{fig:J04_filtering_threshold}) while preserving a sufficient sample size of $\sim$124\,000 pulses (comparable to that available for PSR~J0835$-$4510) for reliable single-pulse analysis.

\subsection{VAE reconstruction and SOM clustering}\label{sec:VAESOM}

Our machine-learning pipeline follows the architecture, latent dimension, and training strategy introduced in \citet{Lousto:2021dia} for both the VAE and the SOM. However, instead of the standard VAE \citep{2013arXiv1312.6114K}, we employ $\beta$-VAE \citep{higgins2017betavae}, which provides a greater control over the balance between reconstruction loss and Kullback--Leibler (KL) divergence by weighting the KL term with a tunable parameter $\beta$, as shown in Eq.~(\ref{eq:beta_vae}). 

\begin{equation} 
\mathcal{L} = \mathbb{E}_{q_{\phi}(\mathbf{y}\mid\mathbf{x})} \left[ \log p_{\theta}(\mathbf{x}\mid\mathbf{y}) \right] - \beta\,D_{\mathrm{KL}} \left( q_{\phi}(\mathbf{y}\mid\mathbf{x}) \,\|\, p(\mathbf{y}) \right) \label{eq:beta_vae}
\end{equation}

The parameters of the VAE encoder and decoder are jointly optimized with the Adam optimizer, with a learning rate of $10^{-4}$. We finetune the VAE hyperparameters separately for each pulsar: for PSR~J0835$-$4510, we set $\beta=0.3$ and train for 30 epochs; for PSR~J1644$-$4559, $\beta=0.1$ with 50 epochs; and for PSR~J0437$-$4715, $\beta=0.3$ with 50 epochs. In addition, the SOM hyperparameters are set to a learning rate of 0.3 and a neighborhood width $\sigma=0.55$, and the maps are fitted using 60\,000 iterations for all pulsars.

After filtering (only for PSR~J0437$-$4715) and outlier removal, the remaining individual pulses in the observation are first reconstructed with the $\beta$-VAE, which acts as a non-linear denoiser: the VAE learns a low-dimensional latent representation of the pulse shapes and reconstructs each pulse from it, effectively separating the coherent signal from the radiometer noise. The reconstructed pulses are then clustered with Self-Organizing Maps, using 4-, 6-, and 9-cluster decompositions to test the robustness of the resulting groupings in PSR~J0835$-$4510 and PSR~J1644$-$4559, and using 5 cluster decompositions in PSR~J0437$-$4715. For PSR~J0835$-$4510 and PSR~J1644$-$4559, clusters \#1 to \#N are labeled in decreasing order of the peak amplitude of their mean reconstructed pulses; for PSR~J0437$-$4715 they are instead labeled in increasing order of peak location since there is little variation in peak amplitude across clusters (see Section~\ref{sec:J04} for details). Cluster \#0 denotes the mean reconstruction over all pulses that remain after filtering and outlier removal. For PSR~J0437$-$4715 we additionally include cluster \#-1 (cluster minus 1) denoting the mean over \textit{all raw pulses} in the observation for comparison with the clusters obtained from the filtered pulses. Moreover, for the analyses presented here, a window of phase bins around the mean peak of each day is selected before training (200 out of 545 bins for Vela, 200 out of 1388 bins for PSR~J1644$-$4559, and 200 out of 281 bins for PSR~J0437$-$4715), and the same 200-phase-bin window is used when computing the quantitative results reported in the tables.

\section{Results}\label{sec:results}

We present the plots of mean cluster reconstructions for all SOM cluster decompositions of PSR~J0835$-$4510, PSR~J1644$-$4559, and PSR~J0437$-$4715 for visual interpretation. Additionally, we report the quantitative characterization per mean cluster reconstructions---all with estimated $1\sigma$ errors---in the tables: the number of pulses in the cluster; the peak location, which is the index of the maximum value of the mean pulse; the peak height, which is the maximum value of the mean pulse; the peak width, which is computed as the full width at half maximum around the peak; and the peak skew, which describes the asymmetry of the mean pulse around the peak. Moreover, to assess the VAE reconstruction fidelity, we report for all pulsars the root mean squared error (RMSE) between the mean VAE reconstruction and the mean raw pulse of each cluster, computed as $\mathrm{RMSE}=[\sum_{i=1}^{N}(\bar{r}_i-\bar{x}_i)^2/N]^{1/2}$, where $\bar{r}_i$ and $\bar{x}_i$ are, respectively, the mean reconstructed and the mean raw pulse in phase bin $i$, and $N=200$ is the number of phase bins considered. For PSR~J0437$-$4715 we additionally report the signal-to-noise ratio (S/N). For each pulse within a cluster, the S/N is calculated as the ratio of the maximum amplitude inside the window to the absolute mean amplitude outside it; the median of these ratios is reported for each cluster. The cluster-specific window is determined following the approach described for the filtering stage of PSR~J0437$-$4715 in Section~\ref{sec:outliers}.

We repeated the experiments on PSR~J0835$-$4510 considering 4-, 6-, and 9-cluster SOM decompositions, in order to compare the results of the wider-band R2 backend with those obtained with the narrower-band ETTUS-A2 configuration in \citet{Lousto:2021dia}. We observe that the 4- and 6-cluster SOM decompositions with R2 have systematic and consistent results across all days, while the 9-cluster decomposition is less stable. Thus, we present the quantitative results in tabular form for the 6-cluster decompositions to provide consistent results with greater granularity in PSR~J0835$-$4510 and PSR~J1644$-$4559. Moreover, in PSR~J0437$-$4715, we consider 5-cluster SOM decomposition for visual and quantitative analyses as it lies between the 4- and 6-cluster settings that proved consistent for the other two pulsars; an odd number of clusters also yields a well-defined central cluster in phase, which can potentially help to determine the time of arrival of the pulses more precisely. The results for each pulsar are presented in Sections~\ref{sec:J08}--\ref{sec:J04}.

\subsection{PSR~J0835$-$4510 (Vela)}\label{sec:J08}

PSR~J0835$-$4510 has a rotation period $P \approx 89.43$~ms at the epoch of our observations, a dispersion measure $\mathrm{DM} \approx 68$~pc~cm$^{-3}$, and a characteristic age of $\sim$11~kyr; at a distance of $\sim$290~pc it is one of the brightest radio pulsars in the sky, with a mean flux density of $\sim$1.05~Jy at 1400~MHz and a pulse width $W_{50} \approx 1.7$~ms \citep[ATNF catalog,][]{Manchester:2004bp}. It undergoes giant glitches quasi-periodically, roughly every two to three years; the latest, its 23rd recorded glitch, occurred on 2024 April 29 (MJD~60429.86961) \citep{Zubieta:2025rud}.

Here we present the pulse-per-pulse analysis of eight observations with R2 bracketing the 2024 glitch: April 17, 18, 19, and 20 (before the glitch) and May 1, 2, 3, and 4 (after the glitch), all of them acquired with a sampling interval of 163.84~$\mu$s (Table~\ref{tab:obslog}). Figures~\ref{fig:Allbg} and \ref{fig:Allag} display the mean reconstructed pulse of each cluster for the 4-, 6-, and 9-cluster SOM decompositions, before and after the glitch, respectively. The corresponding quantitative characterization of the six-cluster decomposition (number of pulses, peak location, peak height, peak width, pulse skewness, and reconstruction RMSE per cluster) is given in Tables~\ref{tab:J08_April} and~\ref{tab:J08_May} in Appendix~\ref{sec:appendixJ08}.

From the 4- and 6-cluster SOM decompositions of each day of observation in Figures~\ref{fig:Allbg} and \ref{fig:Allag}, we can observe a trend of the amplitude decreasing as we move later in the phase; the same pattern is generally observed in 9-cluster SOM decomposition as well but with some inconsistencies. The peak location and peak height in Tables~\ref{tab:J08_April} and~\ref{tab:J08_May} further support this observation. Moreover, the higher amplitude clusters are narrower, with lower width and higher skewness. It is also evident from the clusters that there are fewer pulses in the observation with higher amplitude than lower amplitude. Additionally, as observed in \citet{Lousto:2021dia}, the right shoulder of the average reconstructed pulses for each cluster generally overlap in 4- and 6-cluster SOM decompositions, while this overlap appears less cleanly in 9-cluster setting.

Quantitatively, the phase drift between the brightest and the faintest cluster
of the six-cluster decomposition amounts to 5~bins, i.e. 0.82~ms, on each of the
four days preceding the glitch, and to 5--6~bins on the four days following it,
while the peak height of cluster \#1 exceeds that of cluster \#6 by an order of
magnitude on every day (Tables~\ref{tab:J08_April} and~\ref{tab:J08_May}).
The mean profile itself is remarkably reproducible: over the four pre-glitch days
the width of cluster \#0 is $9.19\pm0.10$~bins ($1.505\pm0.016$~ms, in good
agreement with the catalog $W_{50}=1.7$~ms) and its skewness $3.905\pm0.017$,
i.e. stable at the percent level.

Against this very stable pre-glitch baseline, the four post-glitch days show no
significant change. The width of cluster \#0 on May~1, 2, 3 and~4 is 9.19, 9.28,
8.98 and 9.13~bins, i.e. $9.15\pm0.13$~bins ($1.498\pm0.021$~ms), and its
skewness is $3.900\pm0.025$: relative to the pre-glitch values these are changes
of $-0.5\pm0.9$\% in width and $-0.1\pm0.4$\% in skewness, both consistent with
zero. The phase drift between clusters \#1 and \#6, the amplitude ordering of the
clusters, and the systematic narrowing and increase of skewness with peak height
are likewise unchanged across the glitch. Averaged over all eight days, the width
of the mean profile is $9.17\pm0.11$~bins ($1.502\pm0.018$~ms), reproducible at
the 1\% level.

We stress that this stability is obtained with single pulses extracted using the
timing solution that models the glitch jump together with its two exponential
recovery terms (Section~\ref{sec:IAR}). An error in the folding frequency leaves
each individual pulse untouched but makes the pulse phase drift monotonically
through an observation, smearing any average taken over the whole data set; with
the recovery terms included, the residual frequency error is negligible and no
such smearing is present on any of the eight days. We return to this point, and
to its methodological implications for single-pulse studies around glitches, in
Section~\ref{sec:conclusions}.

\subsection{PSR~J1644$-$4559}\label{sec:J16}

PSR~J1644$-$4559 (B1641$-$45) has a rotation period $P \approx 455$~ms, a high dispersion measure $\mathrm{DM} \approx 478$~pc~cm$^{-3}$, and a characteristic age of $\sim$360~kyr; it is among the brightest pulsars at 1400~MHz, with a mean flux density of $\sim$300~mJy and a pulse width $W_{50} \approx 8$~ms, its profile being strongly scatter-broadened by the ISM \citep[ATNF catalog,][]{Manchester:2004bp}. It is an infrequent glitcher, with four moderate glitches recorded since 1977 \citep{Zubieta:2024ynv}.

\begin{figure*}[!ht]
   \centering
   \includegraphics[width=\textwidth]{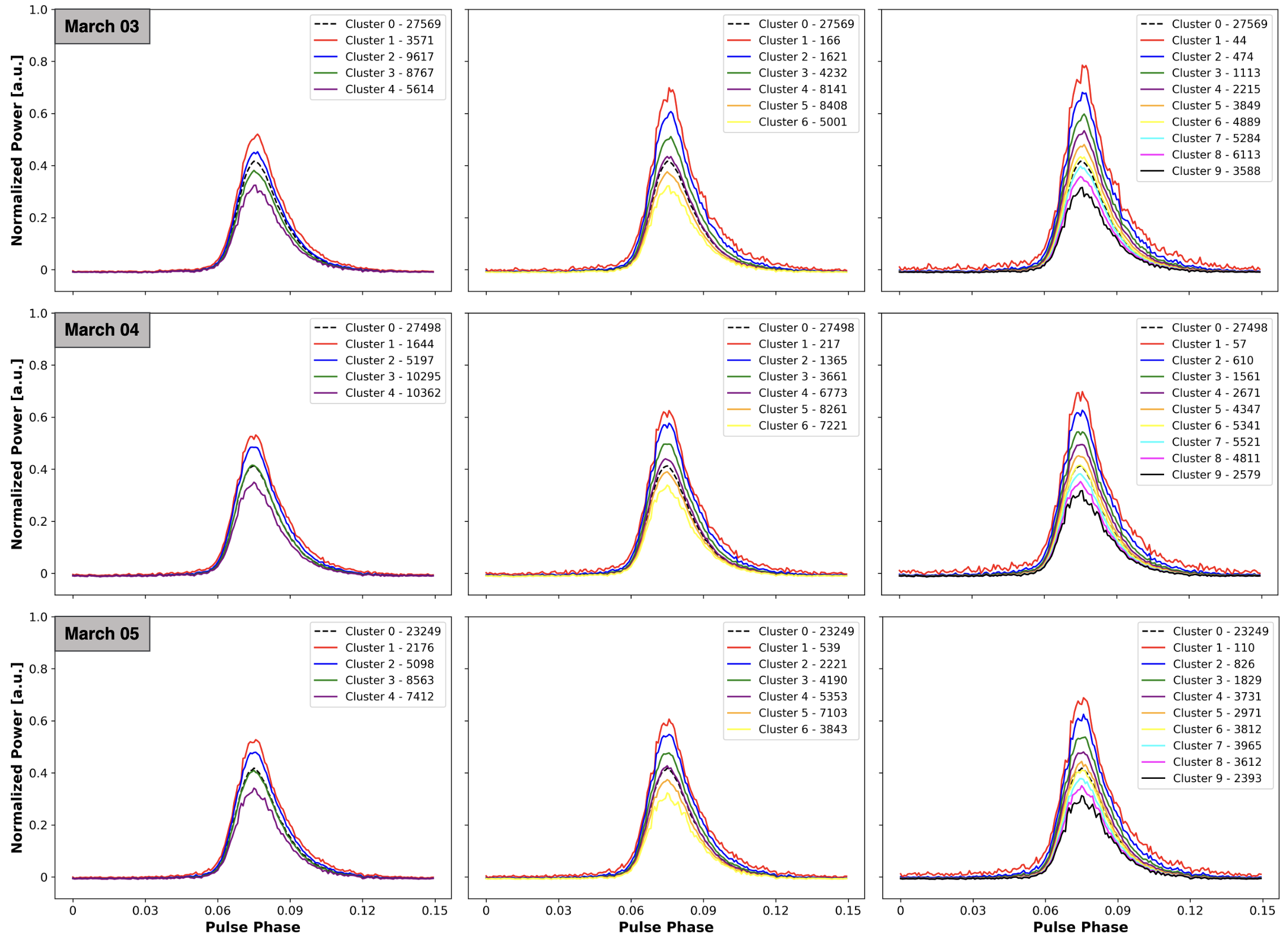}
     \caption{Mean cluster reconstructions for the observations of PSR~J1644$-$4559 with R2 on 2026 March 3 (top row), 4 (middle row), and 5 (bottom row), using 4 (left column), 6 (middle column), and 9 (right column) SOM clusters. The legends give the number of pulses in each cluster. A window of 200 (out of 1388) phase bins around the mean peak of each day is used for the single-pulse analysis and for plotting.}
  \label{fig:J16}
 \end{figure*}

We analyzed three consecutive observations with R2 on 2026 March 3, 4, and 5, acquired with a sampling interval of 327.68~$\mu$s (Table~\ref{tab:obslog}).
Figure~\ref{fig:J16} shows the mean reconstructed pulse per cluster for the 4-, 6-, and 9-cluster SOM decompositions of each day. In Table~\ref{tab:J16} (Appendix~\ref{sec:appendixJ16}) we provide the number of pulses, peak location, peak height, peak width, pulse skewness, and reconstruction RMSE per cluster. All metrics are measured with respect to the 200 phase bins selected out of the 1388 bins per period, each bin corresponding to 327.68~$\mu$s.

In contrast to the amplitude--phase drift observed in PSR~J0835$-$4510 (Vela), the clusters for PSR~J1644$-$4559 vary predominantly in amplitude while exhibiting nearly similar (scattered) shapes. This pattern remains consistent across all three days of observation and within the 4-, 6-, and 9-cluster SOM decompositions for each day (compare Figure~\ref{fig:J16} with Figures~\ref{fig:Allbg}--\ref{fig:Allag}). The peak width and the skewness are very close in value across all clusters, all three SOM configurations (4, 6, and 9 clusters), and all three days of observation (averaging $26.58\pm0.51$~bins, i.e. $8.7\pm0.2$~ms, in peak width, and $1.89\pm0.02$ in peak skew). Additionally, the peak height and peak location of cluster \#0 are nearly constant throughout. The clusters therefore appear to differ only in amplitude.

This behavior is the one expected when the observed pulse shape is set by
propagation rather than by the emission geometry: the scattering tail imposed by
the interstellar medium, of the order of several milliseconds at 1400~MHz, is
common to all pulses and largely erases the intrinsic amplitude--phase
correlation that is so clearly seen in Vela. The constancy of the measured
widths across clusters and days also indicates that we do not resolve
pulse-to-pulse variations of the scattering timescale in the present data set;
setting quantitative limits on such variations will require the polarimetric and
higher-cadence observations discussed in Section~\ref{sec:conclusions}.

\subsection{PSR~J0437$-$4715}\label{sec:J04}

PSR~J0437$-$4715 is the closest and brightest millisecond pulsar known \citep{Johnston1993}, with a rotation period $P \approx 5.757$~ms, $\mathrm{DM} \approx 2.6$~pc~cm$^{-3}$, and a distance of only $\sim$157~pc; it is in a 5.74-day orbit with a helium white-dwarf companion and has a mean flux density of $\sim$150~mJy at 1400~MHz, with a very narrow pulse width of $W_{50} \approx 0.141$~ms \citep[ATNF catalog,][]{Manchester:2004bp}. Its brightness and stability make it a cornerstone of pulsar timing arrays, one of the very few MSPs for which single pulses can be individually detected \citep{Ables1997,Vivekanand1998}, and its timing precision is ultimately limited by pulse-to-pulse jitter, measured at the level of $\sim$48~ns at 1400~MHz for one-hour integrations \citep{Parthasarathy:2021bwl,Lam:2020niv}. Previous IAR observing campaigns of this pulsar are described in \citet{SosaFiscella:2020wmm}.

\begin{figure*}[!ht]
   \centering
   \includegraphics[width=\textwidth]{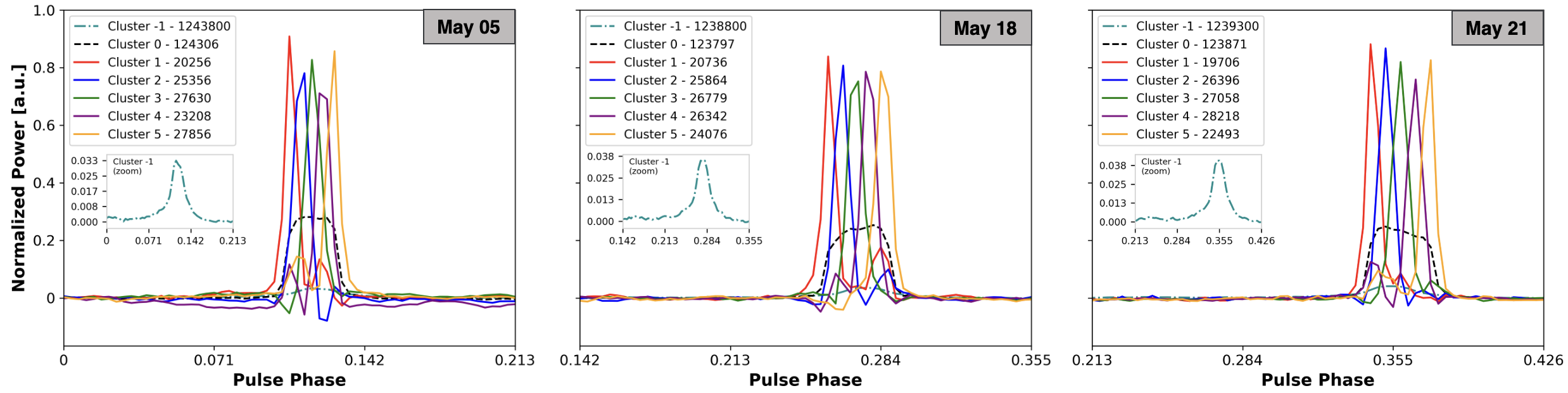}
     \caption{Mean cluster reconstructions for the observations of PSR~J0437$-$4715 with R2 on 2026 May 5 (left column), 18 (middle column), and 21 (right column), using 5 SOM clusters. The legends give the number of pulses in each cluster. The inset plot is a zoomed-in version of Cluster -1, which includes the entire non-filtered raw data set. A window of 200 (out of 281) phase bins around the mean peak of each day is used for the single-pulse analysis, and a zoomed-in version with 60 phase bins is shown in the plots.}
  \label{fig:J04}
 \end{figure*}

We performed a pulse-per-pulse analysis of three observations with R2 on 2026 May 5, 18, and 21. For this millisecond pulsar the sampling interval was reduced to 20.48~$\mu$s (Table~\ref{tab:obslog}), so that its narrow profile is still resolved by several phase bins.
Figure~\ref{fig:J04} shows the mean reconstructed pulse per cluster for the five-cluster SOM decomposition. Table~\ref{tab:J04} (Appendix~\ref{sec:appendixJ04}) presents the number of pulses, peak location, peak height, peak width, pulse skewness, reconstruction RMSE, and S/N for each cluster. All metrics are measured with respect to the 200 phase bins selected out of the 281 bins per period, each bin corresponding to 20.48~$\mu$s.

We can observe in Figure~\ref{fig:J04} that retaining only the pulses with prominent peaks, i.e. going from cluster \#-1 to the filtered sample, improves the signal-to-noise ratio: the S/N of clusters \#0 to \#5 is almost double that of cluster \#-1. Owing to its low S/N, the peak height of cluster \#-1 is significantly lower than those of the other clusters; we have thus provided a zoomed-in inset plot for cluster \#-1 for better visibility. 

Clusters \#1 to \#5 are successively distributed along the pulse phase with slight overlap between the adjoining clusters. These clusters contain comparable numbers of pulses, between $\sim$19\,000 and $\sim$28\,000, and have largely the same peak amplitude. Cluster \#0, representing the mean of all pulses retained after filtering and outlier removal, is 32\% of the average peak height of the clusters \#1 through \#5. Moreover, it spans much of the phase range covered by clusters \#1 through \#5. The five individual clusters are narrower, and have smaller width uncertainties, than either cluster \#0 or cluster \#-1, and this is consistently observed on all three days of observation.

Since the timing precision achievable with a given template scales, to first
order, with the width of the feature being matched, the narrowing of the
individual clusters with respect to the mean profile is the quantity of interest
here. Averaged over the five clusters, the peak width is $2.26$~bins, i.e.
$46$~$\mu$s, to be compared with $8.2$~bins ($168$~$\mu$s) for cluster \#0: the
ratio of the daily means is $3.62\pm0.39$. The $\sim$11\% day-to-day scatter of
this ratio is driven by the width of the mean profile (7.0~bins on May~5 against
8.7--8.9~bins on May~18 and~21) rather than by the cluster widths themselves,
which agree to $\sim$2\% across the three days. We note that cluster \#-1, which contains all the raw pulses, has a width
of $7.4$~bins ($152$~$\mu$s), close to the catalog value
$W_{50}=0.141$~ms for this pulsar, which provides a useful check that our width
estimator recovers the standard profile width when applied to the unfiltered data.

The central cluster \#3 is the natural template for a cluster-based timing
analysis, since its mean pulse is the one aligned with the peak of the integrated
profile (peak locations of 33, 77 and 101~bins on the three days, against 33, 79
and 99 for cluster \#0). Taken individually, its width is smaller than that of
cluster \#0 by a factor $3.28\pm0.40$, slightly below the five-cluster average
because cluster \#3 happens to be the broadest of the five clusters on each of
the three days. Either
estimate suggests that constructing times of arrival from the pulses of a single,
phase-coherent cluster, rather than from the full pulse ensemble, could reduce the
timing uncertainty of PSR~J0437$-$4715 by a factor of $\sim$3.5. We stress that
this is an expectation based on the cluster widths, and not yet a timing
measurement; the corresponding test is deferred to the follow-up work described in
Section~\ref{sec:conclusions}.

\section{Conclusions and discussion}\label{sec:conclusions}

We have presented a homogeneous single-pulse study of three bright southern pulsars observed with the upgraded ROACH-based digital backend (R2) at the IAR, applying the VAE reconstruction and SOM clustering pipeline developed in \citet{Lousto:2021dia}.

For the Vela pulsar, the pulse-per-pulse reanalysis of the eight days bracketing the 2024 April 29 giant glitch confirms, now with R2 data, the results previously obtained with the ETTUS-A1/A2 SDR backends \citep{Lousto:2021dia,Zubieta:2022umm,Zubieta:2025rud}: higher-amplitude pulse clusters systematically peak earlier in phase, display narrower widths, and show larger skewness, consistent with an emission-height stratification within the open field lines. The mean profile is stable at the percent level on all eight days, and the cluster properties reproduce those obtained with the ETTUS receivers; that this is now recovered with a backend of seven times the bandwidth, and with an outlier-rejection stage that removes the spurious clusters which previously contaminated the SOM output, makes the comparison appreciably more robust than in our earlier analyses. Across the glitch itself we find no significant change of the single-pulse organization: with the post-glitch relaxation explicitly included in the timing solution used to extract the pulses, the width of the mean profile agrees before and after the glitch to $-0.5\pm0.9$\%, its skewness to $-0.1\pm0.4$\%, and the amplitude--phase hierarchy of the clusters, including their phase separations, is preserved on every day. This extends to the 2024 event, and sharpens quantitatively, the null results obtained for the 2021 glitch \citep{Zubieta:2022umm} and in our first analysis of the 2024 glitch \citep{Zubieta:2025rud}: on day timescales, the magnetospheric emission of Vela appears remarkably insensitive to even a giant glitch, in contrast with the transient single-pulse phenomenology observed within minutes of the 2016 event \citep{Palfreyman2018}.

This null result carries, in hindsight, a methodological lesson for single-pulse studies around glitches. In an earlier iteration of this analysis, based on a timing solution that absorbed the transient part of the frequency jump into a permanent step, the average post-glitch profile appeared to broaden progressively---by up to 55\% on the fourth day after the glitch---while the individual clusters remained narrow and merely separated in phase. The apparent broadening scaled with the decayed fraction of the transient, corresponding to residual folding-frequency errors of only $\Delta\nu/\nu\sim10^{-7}$, and disappeared entirely once the exponential recovery terms of \citet{Zubieta:2025rud} were incorporated into the \texttt{polyco} models. A folding-frequency error at that level leaves each individual pulse untouched but smears any quantity averaged over a several-hour observation; apparent changes of averaged pulse shapes in the days following a glitch should therefore be validated against the accuracy of the post-glitch timing model before any magnetospheric interpretation is attempted.

The new results for PSR~J1644$-$4559 probe a different type of object: an older, rotationally settled pulsar whose radio emission reaches us strongly scatter-broadened. This is directly reflected in the morphology of its clusters, as shown in Figure~\ref{fig:J16}: instead of the amplitude--phase drift characteristic of Vela, the clusters of PSR~J1644$-$4559 are, to first order, amplitude-scaled versions of a common, scattering-dominated shape, their widths and skewnesses agreeing to within a few percent across all clusters, cluster decompositions, and days. This comparative behavior demonstrates that our pipeline can meaningfully classify single pulses under very different emission and propagation conditions.

The main emphasis of this work is the possibility of improving the timing of the millisecond pulsar PSR~J0437$-$4715. Our analysis shows that selecting the 10\% of pulses with the highest peak-dominance score nearly doubles the signal-to-noise ratio, and that the resulting five SOM clusters are individually narrower than the mean profile by a factor $3.62\pm0.39$ (46~$\mu$s against 168~$\mu$s), which is the figure of merit that ultimately controls the achievable timing precision. This realizes, for this MSP, the prospect anticipated in \citet{Lousto:2023yvu} from the narrower nature of the high-amplitude pulse clusters found in Vela \citep{Lousto:2021dia}. Turning it into an actual improvement requires the construction of a pulsar template \citep{SosaFiscella:2020wmm} to be matched against the observations. Since our single-pulse analysis collects pulses into large clusters and our typical observations extend over $\sim$3 hours, the intrinsic jitter noise floor \citep[$\sim$48~ns at 1400~MHz for one-hour integrations;][]{Parthasarathy:2021bwl,Lam:2020niv} is not currently the main limitation to our precision. 
This improvement is to be confirmed in a follow-up paper using standard timing data and times-of-arrival (TOA) construction, comparing template matching on the full pulse ensemble with TOAs built from the selected high-amplitude clusters. If confirmed, such a cluster-based selection would provide a practical route to mitigate the shape-variability noise of bright MSPs \citep{Oslowski2011,Shannon2014}, of direct relevance to pulsar-timing-array science.

Finally, we note that the continuous, high-cadence monitoring of southern pulsars at IAR is of interest well beyond pulsar astrophysics itself, and the very glitch analyzed in this work provides a concrete illustration. The LVK collaboration has searched the data of its fourth observing run for gravitational waves associated with the 2024 April 29 Vela glitch, both for seconds-long burst-like emission---primarily from fundamental-mode oscillations---and for longer quasi-monochromatic transients of up to four months in duration, primarily from quasi-static quadrupolar deformations \citep{LIGOScientific:2025dup}. No significant candidate was found, but for the first time direct upper limits on the gravitational strain amplitude were set that are stricter than those inferred indirectly from the overall glitch energy scale. Searches of this kind rely on an accurate electromagnetic determination of the glitch epoch and of the post-glitch relaxation---the latter over precisely the weeks-to-months baseline covered by the long-duration search---which is what high-cadence radio monitoring of the kind reported here provides.

More generally, the temporal coincidence of our daily observations with gravitational-wave events detected by the LVK collaboration offers the opportunity to search for correlated electromagnetic signatures---for instance, glitches or emission-state changes contemporaneous with candidate Galactic gravitational-wave transients---as well as to contribute radio monitoring of electromagnetic counterparts identified at other wavelengths. The single-pulse sensitivity demonstrated here, combined with the flexibility of a dedicated facility able to react promptly to alerts, makes the IAR program a valuable component of multimessenger follow-up campaigns in the southern sky.

\begin{acknowledgments}
COL gratefully acknowledges support from NSF awards AST-2319326, PHY-2207920, and PHY-2513442. We are very grateful to the staff of the IAR for their continuous technical support during the execution of this work. S.B.A.F. and E.Z. are PhD candidates with CONICET fellowships. F.G. acknowledges support from PIBAA 1275 and PIP 0113 (CONICET). S.d.P. acknowledges support from ERC Advanced Grant 789410. This research has made use of the ATNF Pulsar Catalogue \citep{Manchester:2004bp}.
\software{PRESTO \citep{presto}, PSRCHIVE \citep{hotan_psrchive_2004},
RFIClean \citep{Maan:2020ent}, scikit-learn \citep{pedregosa2011scikit}, SciPy}
\end{acknowledgments}

\bibliographystyle{aasjournalv7}
\bibliography{PPPPP}

\appendix

\section{Tables of the SOM clustering for the Vela pulsar (PSR~J0835$-$4510)}\label{sec:appendixJ08}

In Tables~\ref{tab:J08_April} and~\ref{tab:J08_May} we describe in detail the six-cluster analysis of the R2 observations taken on 2024 April 17, 18, 19, and 20 (before the glitch), and on 2024 May 1, 2, 3, and 4 (after the glitch).

\begin{figure*}[!ht]
   \centering
   \includegraphics[width=\textwidth]{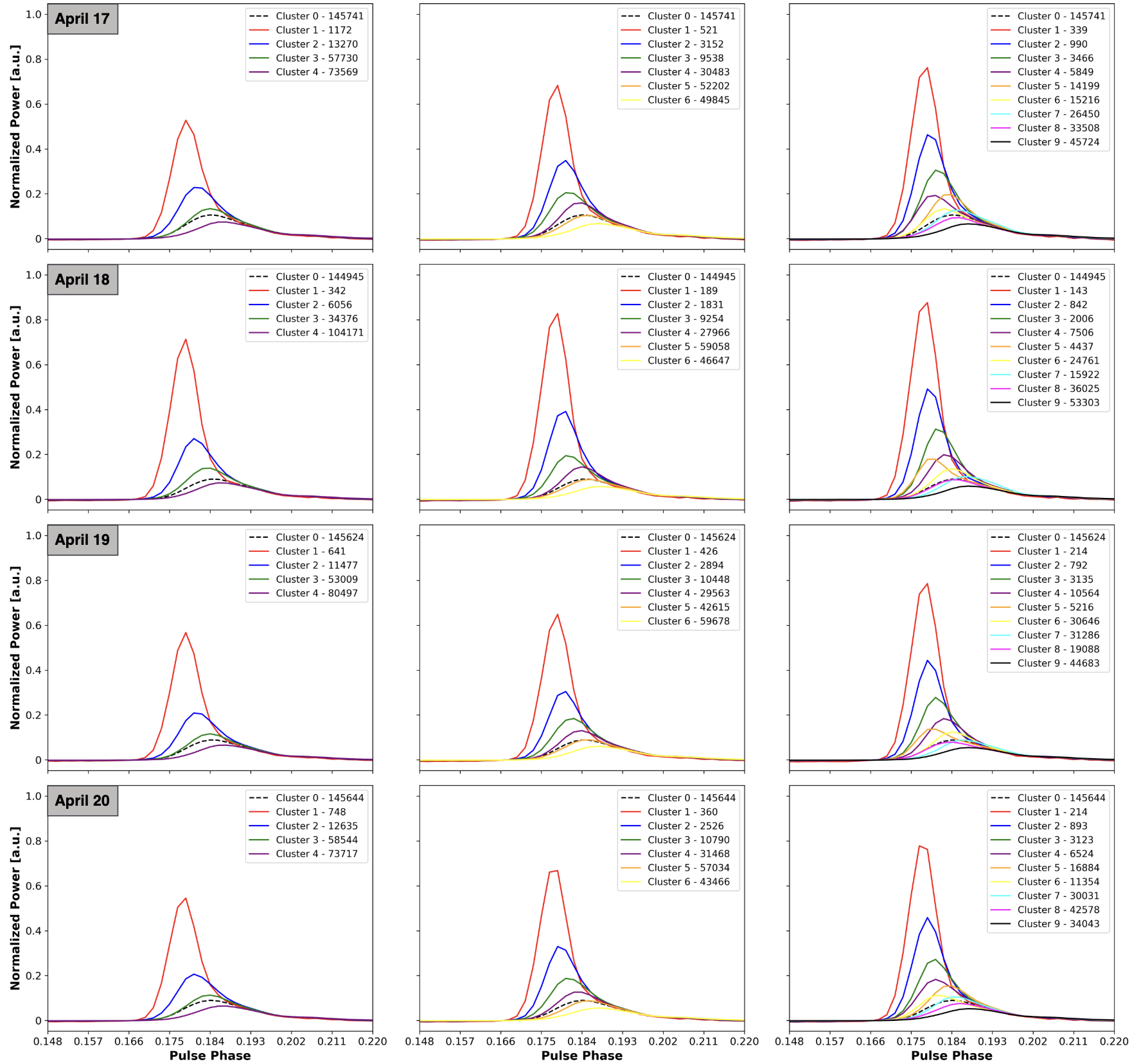}
     \caption{Mean cluster reconstructions for the observations of Vela (PSR~J0835$-$4510) with R2 before the glitch, on 2024 April 17, 18, 19, and 20 (top to bottom), using 4 (left column), 6 (middle column), and 9 (right column) SOM clusters. The legends give the number of pulses in each cluster. A window of 200 (out of 545) phase bins around the mean peak of each day is used for the single-pulse analysis, and a zoomed-in version with 50 phase bins is shown in the plots.}
  \label{fig:Allbg}
 \end{figure*}

\begin{figure*}[!ht]
   \centering
   \includegraphics[width=\textwidth]{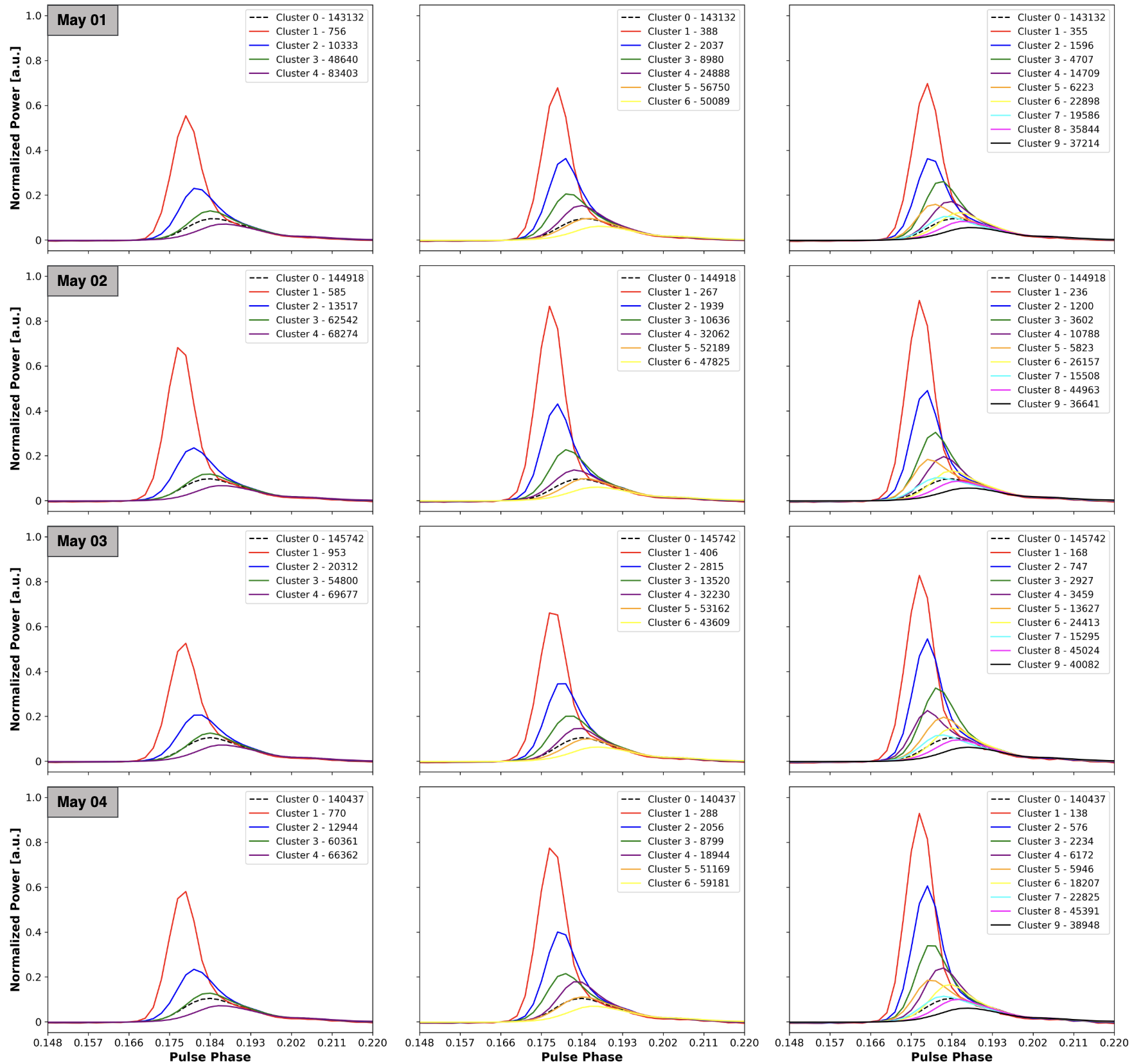}
     \caption{Same as Figure~\ref{fig:Allbg}, but after the glitch, on 2024 May 1, 2, 3, and 4 (top to bottom).}
  \label{fig:Allag}
 \end{figure*}

The columns are the quantities defined in Section~\ref{sec:results}: the number of pulses, and the location, height, width and skewness of the peak of the mean pulse, followed by the reconstruction RMSE. The peak width is the full width at half maximum obtained with \texttt{scipy.signal.peak\_widths}\footnote{\url{https://docs.scipy.org/doc/scipy/reference/generated/scipy.signal.peak_widths.html}} and the skewness is the Fisher--Pearson coefficient from \texttt{scipy.stats.skew}\footnote{\url{https://docs.scipy.org/doc/scipy/reference/generated/scipy.stats.skew.html}}. All metrics carry estimated $1\sigma$ errors and are computed over the 200-phase-bin window of Section~\ref{sec:VAESOM}, taken from the 545 bins into which the period is divided (163.84~$\mu$s per bin). Cluster \#0 is the mean over all pulses surviving outlier rejection; clusters \#1 to \#N are ordered by decreasing peak amplitude of their mean reconstructed pulse, as in Figure~\ref{fig:Allbg} and \ref{fig:Allag}.

We observe a systematic tendency for the pulse peaks to appear earlier in phase the higher the amplitude, together with a reduction of the width and an increase of the skewness, as also found in our previous analyses of the 2021 observations \citep{Lousto:2021dia,Zubieta:2022umm}. Moreover, the peak of cluster \#0 is centered at around bin 100.

\input{J08_April}
\input{J08_May}

Figures~\ref{fig:Allbg} and \ref{fig:Allag} display all the days of observation used in the 4-, 6-, and 9-cluster SOM analyses.

\section{Tables of the SOM clustering for PSR~J1644$-$4559}\label{sec:appendixJ16}

In Table~\ref{tab:J16} we list, for the three
observations of 2026 March 3, 4, and 5, the six-cluster SOM decomposition of
PSR~J1644$-$4559. The columns have the same meaning as in
Appendix~\ref{sec:appendixJ08}; here the metrics are computed over the
200-phase-bin window taken from the 1388 bins per period (327.68~$\mu$s per bin).
Clusters \#1 to \#N are ordered by decreasing peak amplitude of their mean
reconstructed pulse, as in Figure~\ref{fig:J16}, and cluster \#0 is the mean over
all pulses surviving outlier rejection.

\input{J16_March}

\section{Tables of the SOM clustering for PSR~J0437$-$4715}\label{sec:appendixJ04}

In Table~\ref{tab:J04} we list, for the three
observations of 2026 May 5, 18, and 21, the five-cluster SOM decomposition of
PSR~J0437$-$4715, with the addition of the median S/N per cluster defined in
Section~\ref{sec:results}. The metrics are computed over the 200-phase-bin window
taken from the 281 bins per period (20.48~$\mu$s per bin). In contrast to the other
two pulsars, clusters \#1 to \#N are here ordered by increasing peak location,
since their peak amplitudes are nearly identical. With this convention cluster
\#3 is the central one in phase on each of the three days (peak locations 33, 77 and 101~bins, against 33, 79 and 99 for cluster \#0), which is why it is the
reference adopted for the timing discussion of Section~\ref{sec:J04}; cluster \#0 is the mean over
the filtered pulses surviving outlier rejection, and cluster \#-1 the mean over
all raw pulses of the observation. No RMSE is quoted for cluster \#-1, since no
VAE reconstruction is produced for the unfiltered data set.

\input{J04_May}

\end{document}

%% file: J08_April.tex
\begin{table*}
\caption{SOM clustering of PSR~J0835$-$4510 (Vela) with antenna R2 on the four days preceding the 2024 glitch.}
\label{tab:J08_April}
\centering
\begin{tabular}{cclllll}
\hline\hline
Cluster \# & \# Pulses & Peak Loc & Peak Height & Peak Width & Peak Skew & RMSE \\
\hline
\multicolumn{7}{c}{\textbf{2024 April 17}}\\
\hline
0 & 145741 & $100.00 \pm 1.91$ & $22.27 \pm 16.67$ & $9.04 \pm 2.23$ & $3.93 \pm 0.59$ & $0.04$   \\
1 & 521 & $97.00 \pm 0.67$ & $159.30 \pm 46.06$ & $4.18 \pm 0.50$ & $5.96 \pm 0.44$ & $0.50$   \\
2 & 3152 & $98.00 \pm 0.80$ & $73.15 \pm 17.68$ & $5.40 \pm 1.12$ & $5.07 \pm 0.52$ & $0.17$   \\
3 & 9538 & $98.00 \pm 1.03$ & $43.04 \pm 9.86$ & $6.93 \pm 1.68$ & $4.48 \pm 0.49$ & $0.07$   \\
4 & 30483 & $100.00 \pm 1.07$ & $33.51 \pm 7.41$ & $7.23 \pm 1.64$ & $4.37 \pm 0.45$ & $0.06$   \\
5 & 52202 & $101.00 \pm 1.38$ & $21.80 \pm 4.33$ & $8.50 \pm 1.79$ & $4.03 \pm 0.42$ & $0.06$   \\
6 & 49845 & $102.00 \pm 1.44$ & $14.17 \pm 3.65$ & $9.01 \pm 1.72$ & $3.80 \pm 0.38$ & $0.04$   \\
\hline
\multicolumn{7}{c}{\textbf{2024 April 18}}\\
\hline
0 & 144945 & $100.00 \pm 1.98$ & $18.87 \pm 14.88$ & $9.25 \pm 2.20$ & $3.90 \pm 0.57$ & $0.02$   \\
1 & 189 & $97.00 \pm 0.68$ & $193.09 \pm 42.00$ & $4.12 \pm 0.39$ & $6.16 \pm 0.37$ & $0.59$   \\
2 & 1831 & $98.00 \pm 0.78$ & $82.20 \pm 24.24$ & $4.90 \pm 0.87$ & $5.40 \pm 0.54$ & $0.20$   \\
3 & 9254 & $98.00 \pm 0.89$ & $40.82 \pm 10.63$ & $6.41 \pm 1.56$ & $4.59 \pm 0.50$ & $0.08$   \\
4 & 27966 & $100.00 \pm 0.91$ & $30.39 \pm 6.51$ & $7.07 \pm 1.49$ & $4.41 \pm 0.43$ & $0.04$   \\
5 & 59058 & $101.00 \pm 1.38$ & $18.56 \pm 3.67$ & $8.81 \pm 1.75$ & $3.97 \pm 0.39$ & $0.04$   \\
6 & 46647 & $102.00 \pm 1.36$ & $11.89 \pm 2.71$ & $9.16 \pm 1.49$ & $3.77 \pm 0.31$ & $0.03$   \\
\hline
\multicolumn{7}{c}{\textbf{2024 April 19}}\\
\hline
0 & 145624 & $100.00 \pm 1.89$ & $18.66 \pm 14.30$ & $9.22 \pm 2.14$ & $3.90 \pm 0.55$ & $0.03$   \\
1 & 426 & $97.00 \pm 0.63$ & $151.27 \pm 47.55$ & $4.18 \pm 0.35$ & $5.99 \pm 0.37$ & $0.58$   \\
2 & 2894 & $98.00 \pm 0.76$ & $63.96 \pm 16.93$ & $5.25 \pm 1.01$ & $5.13 \pm 0.55$ & $0.23$   \\
3 & 10448 & $99.00 \pm 1.02$ & $38.74 \pm 7.72$ & $6.65 \pm 1.47$ & $4.54 \pm 0.46$ & $0.10$   \\
4 & 29563 & $100.00 \pm 1.09$ & $27.35 \pm 4.96$ & $7.49 \pm 1.53$ & $4.29 \pm 0.41$ & $0.05$   \\
5 & 42615 & $101.00 \pm 1.35$ & $18.79 \pm 3.48$ & $8.72 \pm 1.72$ & $3.98 \pm 0.38$ & $0.04$   \\
6 & 59678 & $102.00 \pm 1.33$ & $12.79 \pm 3.25$ & $9.07 \pm 1.50$ & $3.82 \pm 0.33$ & $0.05$   \\
\hline
\multicolumn{7}{c}{\textbf{2024 April 20}}\\
\hline
0 & 145644 & $100.00 \pm 1.85$ & $18.85 \pm 14.51$ & $9.24 \pm 2.16$ & $3.89 \pm 0.56$ & $0.05$   \\
1 & 360 & $97.00 \pm 0.79$ & $155.80 \pm 50.10$ & $4.29 \pm 0.47$ & $6.02 \pm 0.40$ & $0.92$   \\
2 & 2526 & $97.00 \pm 0.76$ & $69.21 \pm 18.47$ & $5.18 \pm 1.02$ & $5.16 \pm 0.50$ & $0.24$   \\
3 & 10790 & $98.00 \pm 1.00$ & $39.36 \pm 8.61$ & $6.78 \pm 1.53$ & $4.52 \pm 0.47$ & $0.11$   \\
4 & 31468 & $100.00 \pm 1.23$ & $26.45 \pm 5.84$ & $7.73 \pm 1.68$ & $4.25 \pm 0.43$ & $0.07$   \\
5 & 57034 & $101.00 \pm 1.21$ & $18.52 \pm 3.83$ & $8.50 \pm 1.63$ & $4.03 \pm 0.37$ & $0.07$   \\
6 & 43466 & $102.00 \pm 1.40$ & $11.66 \pm 2.48$ & $9.26 \pm 1.53$ & $3.73 \pm 0.31$ & $0.03$   \\
\hline
\end{tabular}
\end{table*}

%% file: J08_May.tex
\begin{table*}
\caption{SOM clustering of PSR~J0835$-$4510 (Vela) with antenna R2 on the four days following the 2024 glitch.}
\label{tab:J08_May}
\centering
\begin{tabular}{cclllll}
\hline\hline
Cluster \# & \# Pulses & Peak Loc & Peak Height & Peak Width & Peak Skew & RMSE \\
\hline
\multicolumn{7}{c}{\textbf{2024 May 1}}\\
\hline
0 & 143132 & $100.00 \pm 1.98$ & $20.01 \pm 15.30$ & $9.19 \pm 2.22$ & $3.90 \pm 0.57$ & $0.08$   \\
1 & 388 & $97.00 \pm 0.68$ & $158.21 \pm 43.41$ & $4.18 \pm 0.46$ & $5.99 \pm 0.45$ & $0.76$   \\
2 & 2037 & $98.00 \pm 0.75$ & $76.30 \pm 17.80$ & $5.11 \pm 1.01$ & $5.22 \pm 0.52$ & $0.23$   \\
3 & 8980 & $98.00 \pm 0.93$ & $43.13 \pm 9.98$ & $6.55 \pm 1.59$ & $4.57 \pm 0.49$ & $0.10$   \\
4 & 24888 & $100.00 \pm 0.95$ & $32.31 \pm 6.46$ & $7.14 \pm 1.52$ & $4.39 \pm 0.43$ & $0.10$   \\
5 & 56750 & $101.00 \pm 1.37$ & $20.28 \pm 4.05$ & $8.61 \pm 1.76$ & $4.00 \pm 0.41$ & $0.09$   \\
6 & 50089 & $102.00 \pm 1.41$ & $12.90 \pm 3.04$ & $9.07 \pm 1.59$ & $3.78 \pm 0.33$ & $0.06$   \\
\hline
\multicolumn{7}{c}{\textbf{2024 May 2}}\\
\hline
0 & 144918 & $100.00 \pm 1.89$ & $20.43 \pm 17.05$ & $9.28 \pm 2.31$ & $3.87 \pm 0.60$ & $0.05$   \\
1 & 267 & $96.00 \pm 0.61$ & $202.06 \pm 44.52$ & $4.05 \pm 0.35$ & $6.20 \pm 0.36$ & $0.57$   \\
2 & 1939 & $97.00 \pm 0.68$ & $90.48 \pm 26.69$ & $4.73 \pm 0.94$ & $5.43 \pm 0.56$ & $0.24$   \\
3 & 10636 & $98.00 \pm 0.95$ & $47.69 \pm 11.45$ & $6.24 \pm 1.45$ & $4.67 \pm 0.51$ & $0.08$   \\
4 & 32062 & $99.00 \pm 1.26$ & $28.74 \pm 6.83$ & $7.91 \pm 1.90$ & $4.17 \pm 0.48$ & $0.07$   \\
5 & 52189 & $100.00 \pm 1.36$ & $20.54 \pm 4.56$ & $8.41 \pm 1.71$ & $4.06 \pm 0.41$ & $0.07$   \\
6 & 47825 & $102.00 \pm 1.46$ & $12.61 \pm 2.95$ & $9.21 \pm 1.66$ & $3.74 \pm 0.34$ & $0.04$   \\
\hline
\multicolumn{7}{c}{\textbf{2024 May 3}}\\
\hline
0 & 145742 & $100.00 \pm 1.86$ & $22.11 \pm 16.06$ & $8.98 \pm 2.18$ & $3.93 \pm 0.58$ & $0.06$   \\
1 & 406 & $96.00 \pm 0.84$ & $154.11 \pm 44.82$ & $4.30 \pm 0.56$ & $5.96 \pm 0.47$ & $1.46$   \\
2 & 2815 & $98.00 \pm 0.94$ & $72.55 \pm 20.97$ & $5.50 \pm 1.23$ & $5.08 \pm 0.57$ & $0.41$   \\
3 & 13520 & $99.00 \pm 1.22$ & $42.14 \pm 10.25$ & $7.04 \pm 1.71$ & $4.47 \pm 0.50$ & $0.13$   \\
4 & 32230 & $100.00 \pm 1.07$ & $30.82 \pm 6.32$ & $7.32 \pm 1.58$ & $4.35 \pm 0.43$ & $0.06$   \\
5 & 53162 & $101.00 \pm 1.37$ & $20.99 \pm 4.26$ & $8.54 \pm 1.75$ & $4.02 \pm 0.41$ & $0.06$   \\
6 & 43609 & $102.00 \pm 1.52$ & $13.29 \pm 3.23$ & $9.21 \pm 1.74$ & $3.74 \pm 0.36$ & $0.04$   \\
\hline
\multicolumn{7}{c}{\textbf{2024 May 4}}\\
\hline
0 & 140437 & $100.00 \pm 1.84$ & $21.95 \pm 16.35$ & $9.13 \pm 2.21$ & $3.90 \pm 0.58$ & $0.07$   \\
1 & 288 & $96.00 \pm 0.72$ & $180.59 \pm 45.91$ & $4.22 \pm 0.44$ & $6.08 \pm 0.39$ & $0.53$   \\
2 & 2056 & $97.00 \pm 0.80$ & $84.10 \pm 23.22$ & $5.07 \pm 0.95$ & $5.27 \pm 0.53$ & $0.27$   \\
3 & 8799 & $98.00 \pm 0.87$ & $45.16 \pm 10.83$ & $6.54 \pm 1.59$ & $4.53 \pm 0.48$ & $0.12$   \\
4 & 18944 & $99.00 \pm 0.91$ & $37.53 \pm 7.42$ & $6.81 \pm 1.40$ & $4.52 \pm 0.42$ & $0.15$   \\
5 & 51169 & $100.00 \pm 1.44$ & $23.35 \pm 4.52$ & $8.77 \pm 1.86$ & $3.99 \pm 0.43$ & $0.08$   \\
6 & 59181 & $101.00 \pm 1.58$ & $14.54 \pm 3.90$ & $9.17 \pm 1.79$ & $3.78 \pm 0.38$ & $0.05$   \\
\hline
\end{tabular}
\end{table*}

%% file: J16_March.tex
\begin{table*}
\caption{SOM clustering of PSR~J1644$-$4559 with antenna R2 on 2026 March 3, 4, and 5.}
\label{tab:J16}
\centering
\begin{tabular}{cclllll}
\hline\hline
Cluster \# & \# Pulses & Peak Loc & Peak Height & Peak Width & Peak Skew & RMSE \\
\hline
\multicolumn{7}{c}{\textbf{2026 March 3}}\\
\hline
0 & 27569 & $100.00 \pm 1.91$ & $2.50 \pm 0.57$ & $26.67 \pm 3.97$ & $1.91 \pm 0.06$ & $0.01$   \\
1 & 166 & $101.00 \pm 1.80$ & $4.66 \pm 0.72$ & $26.17 \pm 3.73$ & $1.89 \pm 0.05$ & $0.21$   \\
2 & 1621 & $102.00 \pm 1.82$ & $3.65 \pm 0.36$ & $26.77 \pm 3.56$ & $1.89 \pm 0.05$ & $0.07$   \\
3 & 4232 & $102.00 \pm 1.97$ & $3.07 \pm 0.26$ & $26.65 \pm 3.31$ & $1.90 \pm 0.05$ & $0.03$   \\
4 & 8141 & $100.00 \pm 1.83$ & $2.61 \pm 0.25$ & $26.48 \pm 3.22$ & $1.91 \pm 0.05$ & $0.02$   \\
5 & 8408 & $100.00 \pm 1.84$ & $2.26 \pm 0.21$ & $26.10 \pm 3.54$ & $1.92 \pm 0.06$ & $0.02$   \\
6 & 5001 & $101.00 \pm 1.85$ & $1.93 \pm 0.20$ & $26.76 \pm 5.36$ & $1.89 \pm 0.08$ & $0.04$   \\
\hline
\multicolumn{7}{c}{\textbf{2026 March 4}}\\
\hline
0 & 27498 & $100.00 \pm 1.82$ & $2.48 \pm 0.52$ & $26.57 \pm 3.86$ & $1.90 \pm 0.06$ & $0.02$   \\
1 & 217 & $101.00 \pm 2.13$ & $4.17 \pm 0.60$ & $26.16 \pm 3.76$ & $1.87 \pm 0.05$ & $0.19$   \\
2 & 1365 & $101.00 \pm 2.03$ & $3.46 \pm 0.36$ & $26.54 \pm 3.46$ & $1.89 \pm 0.05$ & $0.07$   \\
3 & 3661 & $101.00 \pm 2.04$ & $2.98 \pm 0.30$ & $26.94 \pm 3.64$ & $1.90 \pm 0.05$ & $0.04$   \\
4 & 6773 & $99.00 \pm 1.89$ & $2.64 \pm 0.25$ & $26.69 \pm 3.51$ & $1.91 \pm 0.05$ & $0.02$   \\
5 & 8261 & $100.00 \pm 1.77$ & $2.34 \pm 0.21$ & $26.32 \pm 3.47$ & $1.92 \pm 0.05$ & $0.02$   \\
6 & 7221 & $100.00 \pm 1.99$ & $2.03 \pm 0.20$ & $26.28 \pm 4.45$ & $1.89 \pm 0.07$ & $0.04$   \\
\hline
\multicolumn{7}{c}{\textbf{2026 March 5}}\\
\hline
0 & 23249 & $100.00 \pm 2.01$ & $2.51 \pm 0.58$ & $27.22 \pm 5.09$ & $1.88 \pm 0.06$ & $0.02$   \\
1 & 539 & $101.00 \pm 2.13$ & $4.04 \pm 0.56$ & $26.22 \pm 4.01$ & $1.87 \pm 0.05$ & $0.15$   \\
2 & 2221 & $101.00 \pm 2.06$ & $3.29 \pm 0.37$ & $27.54 \pm 4.62$ & $1.88 \pm 0.05$ & $0.05$   \\
3 & 4190 & $101.00 \pm 2.13$ & $2.86 \pm 0.31$ & $27.37 \pm 4.37$ & $1.89 \pm 0.05$ & $0.03$   \\
4 & 5353 & $100.00 \pm 1.86$ & $2.56 \pm 0.25$ & $26.93 \pm 4.04$ & $1.89 \pm 0.05$ & $0.03$   \\
5 & 7103 & $100.00 \pm 1.98$ & $2.24 \pm 0.25$ & $26.72 \pm 4.77$ & $1.89 \pm 0.06$ & $0.03$   \\
6 & 3843 & $100.00 \pm 2.39$ & $1.94 \pm 0.22$ & $26.29 \pm 6.14$ & $1.85 \pm 0.08$ & $0.04$   \\
\hline
\end{tabular}
\end{table*}

%% file: J04_May.tex
\begin{table*}
\caption{SOM clustering of PSR~J0437$-$4715 with antenna R2 on 2026 May 5, 18, and 21.}
\label{tab:J04}
\centering
\begin{tabular}{ccllllll}
\hline\hline
Cluster \# & \# Pulses & Peak Loc & Peak Height & Peak Width & Peak Skew & RMSE & SNR \\
\hline
\multicolumn{8}{c}{\textbf{2026 May 5}}\\
\hline
-1 & 1243800 & $33.00 \pm 87.49$ & $0.40 \pm 16.07$ & $6.94 \pm 4.00$ & $3.73 \pm 3.72$ & - &$36.63$   \\
0 & 124306 & $33.00 \pm 2.02$ & $3.37 \pm 7.95$ & $7.02 \pm 4.81$ & $4.99 \pm 2.00$ & $0.03$ & $74.15$   \\
1 & 20256 & $30.00 \pm 0.81$ & $10.90 \pm 2.46$ & $1.87 \pm 0.59$ & $9.38 \pm 3.68$ & $0.05$ & $67.69$   \\
2 & 25356 & $32.00 \pm 0.71$ & $9.37 \pm 2.87$ & $2.44 \pm 0.68$ & $8.50 \pm 2.96$ & $0.07$ & $71.62$   \\
3 & 27630 & $33.00 \pm 0.72$ & $9.92 \pm 2.38$ & $2.49 \pm 0.82$ & $8.26 \pm 3.01$ & $0.05$ & $71.04$   \\
4 & 23208 & $34.00 \pm 1.18$ & $8.53 \pm 3.09$ & $2.36 \pm 0.70$ & $8.25 \pm 3.06$ & $0.05$ & $67.89$   \\
5 & 27856 & $36.00 \pm 1.81$ & $10.28 \pm 2.45$ & $1.91 \pm 0.66$ & $9.05 \pm 3.69$ & $0.05$ & $66.13$   \\
\hline
\multicolumn{8}{c}{\textbf{2026 May 18}}\\
\hline
-1 & 1238800 & $77.00 \pm 61.63$ & $0.42 \pm 16.13$ & $7.57 \pm 4.55$ & $3.79 \pm 3.77$ & - &$40.31$   \\
0 & 123797 & $79.00 \pm 3.16$ & $3.04 \pm 9.57$ & $8.69 \pm 6.67$ & $4.44 \pm 3.20$ & $0.04$ & $75.96$   \\
1 & 20736 & $73.00 \pm 3.21$ & $10.06 \pm 3.05$ & $1.87 \pm 0.53$ & $8.91 \pm 2.75$ & $0.07$ & $66.01$   \\
2 & 25864 & $75.00 \pm 1.88$ & $9.69 \pm 3.42$ & $2.29 \pm 0.77$ & $8.68 \pm 2.49$ & $0.05$ & $72.67$   \\
3 & 26779 & $77.00 \pm 1.04$ & $9.03 \pm 4.05$ & $2.50 \pm 0.97$ & $8.51 \pm 2.36$ & $0.06$ & $72.90$   \\
4 & 26342 & $78.00 \pm 1.11$ & $9.43 \pm 3.94$ & $2.36 \pm 0.89$ & $8.74 \pm 2.39$ & $0.05$ & $72.12$   \\
5 & 24076 & $80.00 \pm 0.95$ & $9.44 \pm 4.06$ & $2.46 \pm 0.84$ & $8.51 \pm 2.38$ & $0.05$ & $74.55$   \\
\hline
\multicolumn{8}{c}{\textbf{2026 May 21}}\\
\hline
-1 & 1239300 & $100.00 \pm 57.31$ & $0.49 \pm 16.02$ & $7.74 \pm 4.69$ & $3.79 \pm 3.77$ & - &$40.85$   \\
0 & 123871 & $99.00 \pm 3.15$ & $2.97 \pm 9.42$ & $8.89 \pm 6.85$ & $4.36 \pm 3.19$ & $0.04$ & $72.74$   \\
1 & 19706 & $97.00 \pm 1.40$ & $10.58 \pm 3.25$ & $2.04 \pm 0.72$ & $9.34 \pm 2.97$ & $0.06$ & $70.37$   \\
2 & 26396 & $99.00 \pm 1.06$ & $10.40 \pm 3.07$ & $2.33 \pm 0.76$ & $8.78 \pm 2.46$ & $0.05$ & $74.46$   \\
3 & 27058 & $101.00 \pm 0.89$ & $9.83 \pm 3.32$ & $2.51 \pm 1.00$ & $8.32 \pm 2.09$ & $0.05$ & $74.87$   \\
4 & 28218 & $103.00 \pm 1.96$ & $9.10 \pm 3.31$ & $2.44 \pm 0.87$ & $8.25 \pm 2.38$ & $0.07$ & $72.74$   \\
5 & 22493 & $105.00 \pm 2.34$ & $9.91 \pm 3.22$ & $2.06 \pm 0.64$ & $9.19 \pm 3.11$ & $0.05$ & $70.86$   \\
\hline
\end{tabular}
\end{table*}